\documentclass[11pt]{article}

\usepackage{graphicx}  
\usepackage{dcolumn}   
\usepackage{bm}        
\usepackage[english]{babel}
\usepackage{amsfonts,amsmath,amssymb,mathrsfs}
\usepackage{epstopdf}
\usepackage{subfigure}
\usepackage{color}

\def\a{\alpha}
\def\r{\rho}
\def\s{\sigma}
\def\t{\tau}
\def\m{\mu}
\def\n{\nu}
\def\k{\kappa}
\def\th{\theta}
\def\g{\gamma}\def\G{\Gamma}
\def\L{\Lambda}\def\l{\lambda}
\def\D{\Delta}
\def\la{\langle}
\def\ra{\rangle}
\def\o{\omega}\def\O{\Omega}
\def\d{\delta}
\def\p{\partial}

\def\half{\textstyle{\frac{1}{2}}}

\def\bdoc{\begin{document}}
\def\edoc{\end{document}}

\def\beq{\begin{equation}}
\def\eeq{\end{equation}}
\def\bea{\begin{eqnarray}}
\def\eea{\end{eqnarray}}
\def\ben{\begin{enumerate}}
\def\een{\end{enumerate}}
\def\la{\langle}
\def\ra{\rangle}
\def\a{\alpha}
\def\b{\beta}
\def\g{\gamma}\def\G{\Gamma}
\def\d{\delta}\def\D{\Delta}
\def\e{\epsilon}
\def\z{\zeta}

\def\th{\theta}
\def\k{\kappa}
\def\l{\lambda}
\def\m{\mu}
\def\n{\nu}
\def\o{\omega}
\def\p{\pi}
\def\r{\rho}
\def\s{\sigma}
\def\t{\tau}
\def\L{${\cal L}$}
\def\H{${\cal H}$}
\def\S{\Sigma }
\def\gsim{\; \raisebox{-.8ex}{$\stackrel{\textstyle >}{\sim}$}\;}
\def\lsim{\; \raisebox{-.8ex}{$\stackrel{\textstyle <}{\sim}$}\;}
\def\gtrsim{\gsim}
\def\lessim{\lsim}
\def\loc{{\rm local}}
\def\vm{v_{\rm max}}
\def\bh{\bar{h}}
\def\del{\partial}
\def\nab{\nabla}
\def\half{{\textstyle{\frac{1}{2}}}}
\def\fourth{{\textstyle{\frac{1}{4}}}}

\def\bD{{\bf D}}
\def\bE{{\bf E}}
\def\bF{{\bf F}}
\def\bB{{\bf B}}
\def\bP{{\bf P}}
\def\bV{{\bf v}}
\def\bv{{\bf v}}
\def\bx{{\bf x}}
\def\by{{\bf y}}
\def\bz{{\bf z}}
\def\ba{{\bf a}}
\def\bd{{\bf d}}
\def\bs{{\bf s}}
\def\bn{{\bf n}}
\def\bp{{\bf p}}

\def\O{\Omega}

\def\del{\nabla}
\def\br{{\bf r}}
\def\bnab{{\bf \nab}}
\def\lf{\left (}
\def\rt{\right)}
\def\tE{\tilde{E}}
\def\tL{\tilde{L}}
\def\Horava{Ho\v{r}ava }
\def\le{\left }
\def\re{\right}
\begin{document}
\title{Second Law Violations in Lovelock Gravity for Black hole Mergers} 
\author{
Sudipta Sarkar\footnote{sudipta@umd.edu}\vspace*{0.1in}~ and
Aron C. Wall \vspace*{0.1in} \footnote{aronwall@umd.edu}\\
{\footnotesize \it Maryland Center for Fundamental Physics, Department of Physics, University of Maryland }\\{\footnotesize \it  College Park, MD 20742-4111. USA}\vspace*{0.1in} \\ 
}
\date{September 16, 2011}
\maketitle
\begin{abstract}
We study the classical second law of black hole thermodynamics, for Lovelock theories (other than General Relativity), in arbitrary dimensions.  Using the standard formula for black hole entropy, we construct scenarios involving the merger of two black holes in which the entropy instantaneously decreases.  Our construction involves a Kaluza-Klein compactification down to a dimension in which one of the Lovelock terms is topological.  We discuss some open issues in the definition of the second law which might be used to remove this entropy decrease.
\end{abstract}
\section{Introduction}

The basic motivation for studying black hole thermodynamics is to gain insight into the quantum nature of gravity.  Irrespective of the true microscopic theory of quantum gravity, one expects semiclassical gravity to be a low energy effective description.  On general grounds, the action of such an effective theory should consist of classical Einstein-Hilbert action plus a series of additional covariant, higher order terms.  These terms typically arise due to quantum renormalization effects.  However, it is also possible to add them to the action classically, in order to study their effect on classical black hole thermodynamics.\footnote{In string theory, there are also string corrections to the Einstein-Hilbert action at $\mathcal{O}(\hbar^0)$}  Such a study might impose restrictions on the form of these classical terms---if so, it would suggest that horizon statistical mechanics somehow places nontrivial constraints on the form of the low energy action.


In classical General Relativity, the analogue of the second law of thermodynamics is the ``area theorem", which asserts that the area ${\cal A}$ of a black hole can not decrease in any classical process \cite{hawking, ellis}.  Once quantum effects are included, the classical version of the second law is replaced by a generalized version \cite{bekenstein}, saying that ${\cal A}/4G \hbar$ plus the exterior matter entropy is non-decreasing. However, in this work, we will consider only the classical second law neglecting all the quantum effects of the matter fields outside the horizon.


The area theorem in General Relativity depends on the null convergence condition: $R_{ij} k^i k^j > 0 $, for any null vector $k^i$.  For General Relativity, this condition is implied by Einstein's equation together with the null energy condition.  Once one ventures beyond General Relativity, the Einstein equation has corrections, so it is not possible to establish an area increase theorem for such theories.  However, one also needs to add corrections to the Bekenstein formula ${\cal A}/4G \hbar$ in order to get an entropy which satisfies the ``first law'' (really the Clausius relation) $dE = T dS$.  One expects that these two corrections combine to produce a classical second law in terms of the corrected entropy.  For example, this is what happens in $f(R)$ gravity \cite{tedkang}.


The Wald Noether charge method \cite{wald1,wald} can be used to derive a horizon entropy $S$ satisfying the first law in any classical diffeomorphism invariant theory of gravity.  In this approach, the entropy can be expressed as a local geometrical quantity $Q$ integrated over a spacelike cross-section of a Killing horizon.  The integral of $Q$ is the Noether charge of diffeomorphisms under the Killing field, which is a boundary term.  In general, this Noether charge has several ambiguities \cite{wald,tedon}, but for a stationary horizon with a regular bifurcation surface, none of these ambiguities matter and Wald's method uniquely prescribes the black hole entropy.  This entropy $S$ then satisfies the first law for any first order perturbation to the spacetime.  However, any proof of classical second law requires going beyond the stationary setting and analyzing a truely dynamical situation.  Then the ambiguity in defining $S$ matters, and presumably at most one of the possible choices of $S$---the one that actually corresponds to the entropy---will obey a second law.


In this paper, we will analyze the second law for Lovelock gravity, perhaps the next simplest classical gravity theory after $f(R)$, for which no second law has yet been established.  There is a particularly simple way to resolve the Wald ambiguities in this theory in order to obtain an entropy for black holes (often uncritically accepted as \emph{the} Wald entropy for Lovelock black holes).  We will establish that for any Lovelock theory (except General Relativity), the classical second law can be violated in certain black hole mergers.

In the special case of $D = 4$,  this second law violation is already well known \cite{ted, Liko}.  In this case, the only additional Lovelock term in the action (apart from the familiar Einstein-Hilbert term) is topological, so the equation of motion is still the Einstein equation.  Yet the entropy has a correction term which is proportional to the Euler number of the horizon, and this can lead to violations of the second law when the horizon topology changes through black hole collapse or merger.  However, since the theory obeys the Einstein equation, it actually \emph{does} obey a second law, if the entropy is taken to be the area instead of the Wald entropy.  One possible viewpoint is that there is an additional ``topological'' ambiguity in the entropy $S$ in addition to those ambiguities identified in Refs. \cite{tedon, wald}.  Since first order perturbations to a black hole cannot change the topology, the addition of a topological term to $S$ cannot change whether the first law is satisfied or not.  A similar analysis can be applied to higher order Lovelock terms which are topological in $D = 2n > 4$ dimensions.

However, in $D > 4$, the Gauss-Bonnet action is not topological, and therefore can not be viewed as an ambiguity in the entropy for first law purposes.  We will show that even in this non-topological case, the second law must be violated. The conceptual problem must therefore be resolved in some other way.

The outline of the paper is as follows:  In section \ref{love} we describe the Lovelock action.  In section \ref{top} we discuss violations of the second law in topological Lovelock theories.   In section \ref{GB} we consider Gauss-Bonnet theory in $D = 5$, which can be shown to have a second law violation by means of a Kaluza-Klein dimensional reduction to a $D = 4$ theory.  In section \ref{arb} we extend this argument to arbitrary Lovelock theories, and explain why the argument does not apply to General Relativity.  In section \ref{dis} we discuss possible ways to try to reconcile this result with black hole thermodynamics.


Henceforth, we adopt natural units ($G = c = \hbar = 1$)

\section{Lovelock gravity}\label{love}
 A natural generalization of the Einstein-Hilbert Lagrangian is provided by the Lanczos-Lovelock Lagrangian \cite{lovelock}, which is the sum of dimensionally extended Euler densities:
\bea
{\cal L}^{(D)} = \sum_{m=1}^{K} c_m {\cal L}_{m}^{(D)},
\eea
where $c_m$'s are arbitrary constants and ${\cal L}_{m}^{(D)}$ is the $m$-th order Lovelock term, given by
\bea
{\cal L}_{m}^{(D)} = 2^{-m} \delta^{i_1 j_1 ... i_m j_m}_{k_1 l_1 ...k_m l_m} R^{k_1 l_1}_{i_1 j_1}...R^{k_m l_m}_{i_m j_m},
\eea
where $R^{i j}_{k l}$ is the $D$ dimensional curvature tensor and the tensor $\delta^{...}_{...}$ is anti-symmetric in both sets of indices. For $D = 2m$, ${\cal L}_{m}^{(D)}$ is the Euler density of the $2m$ dimensional manifold.  The Einstein-Hilbert action is the specific case in which all the coefficients except $c_1$ are zero.  The most important property of these Lanzcos-Lovelock Lagrangians is that they give second order field equations.  Also, these Lagrangians are free from ghosts when expanded around flat spacetime \cite{string1}.

We would like to first concentrate on the case $m=2$, which in a general $D$ dimensional spacetime, is the action functional
\bea
{\cal I}= \frac{1}{16 \pi} \int d^D x \sqrt{-g} \left( R + \alpha  {\cal L}_{GB}  \right) \label{actionGB},
\eea
where $R$ is the $D$ dimensional Ricci scalar, and $ {\cal L}_{GB}$ is the Gauss-Bonnet invariant of the form
\bea
 {\cal L}_{GB} = R^2 - 4 R_{i j} R^{i j} + R_{ijkl}R^{ijkl}.
\eea


 
For any diffeomorphism invariant theory of gravity described by a Lagrangian ${\cal L}$, the entropy of a stationary black hole with a regular bifurcation surface is given by Ref. \cite{tedon,wald}
 \bea
S = - \frac{1}{8} \int_{{\cal B}} \frac{\partial {\cal L}} {\partial R_{ijkl}} \epsilon_{ij} \epsilon_{kl} ~\sqrt{\sigma}~ d^{D-2} x, \label{entropyWald}
 \eea
where the integration is over any $(D-2)$-dimensional spacelike cross-section ${\cal B}$ of the horizon, and $\epsilon_{ij}$ is the binormal on such a cross-section. The binormals are normalized as, $\epsilon_{ij} ~\epsilon^{ij} = -2$.  For the action in Eq.~(\ref{actionGB}), the entropy turns out to be \cite{ted}
 \bea
 S = \frac{1}{4} \int_{{\cal B}} \left( 1 + 2 \alpha ~^{(D-2)}R \right)~ \sqrt{\sigma}~d^{D-2} x \label{entropyGB},
 \eea
where $^{(D-2)}R $ is the Ricci curvature associated with the $(D-2)$-dimensional cross-section of the horizon.  An important reminder is the fact that the derivation of Eq.~(\ref{entropyGB}) is crucially dependent on the existence of a stationary Killing field.  
The two expressions Eq.~(\ref{entropyWald}) and Eq.~(\ref{entropyGB}) differ by Wald ambiguity terms which are quadratic in the extrinsic curvature of the horizon, and which vanish for stationary black holes.  For nonstationary black holes, it is therefore unclear whether to use Eq.~(\ref{entropyWald}), Eq.~(\ref{entropyGB}), or some other choice of entropy.  The entropy defined by Eq.~(\ref{entropyGB}) has the advantage that it is simpler, since it depends only on the metric of the horizon.  The Gauss-Bonnet contribution to the entropy is also topological in $D = 4$.  However, we shall show below that this choice of entropy does not obey the second law when black holes merge.  (Eq.~(\ref{entropyWald}) does not obey a second law either; cf. section \ref{dis}.) 

We would like to investigate whether this expression for the black hole entropy obeys an increase theorem, like the area theorem in General Relativity. An ideal and straightforward approach would be to directly compute the change of this entropy along the congruence which generates the horizon. In case of General Relativity, such a calculation is much simpler due to the availability of null focussing equation and the field equation $R_{ij} k^i k^j = 8 \pi T_{ij} k^i k^j$. Since for Gauss-Bonnet gravity, the entropy is no longer proportional to the horizon area, but depends on the curvature of the cross-sections, the null focussing equation is not helpful to study the evolution. As a result, instead of following a direct approach, we take a different path, by constructing a black hole merger situation where this entropy function in Eq.~(\ref{entropyGB}) is shown to decrease.  That will serve as a counter example to the validity of classical second law with this entropy.


\section{Second Law Violation in Topological Theories}\label{top}

In four dimensions, the Gauss-Bonnet invariant is topological and does not affect the equation of motion.  The entropy of the two-dimensional horizon cross-sections can be obtained from Eq.~(\ref{entropyGB}) by the Gauss-Bonnet theorem:
\bea
S = \frac{{\cal A}}{4}  + 4 \pi \alpha \chi,
\eea
where $\chi$ is the Euler number of the horizon slice (for spherical topology $\chi = 2$).  Now in four dimensions, the Gauss-Bonnet term in the action has no effect on the equation of motion, so $\alpha$ can be either negative or positive.  On the other hand, in five dimensions, when the Gauss-Bonnet term has non-trivial contribution, the negative sign for $\alpha$ leads to instability and naked singularities \cite{instability}.  Thus we will mainly focus on the case $\alpha > 0$ even in four dimensions, since ultimately we are interested in applying the conclusions obtained from the four dimensional set-up to a five dimensional spacetime via dimensional reduction.  However, a negative sign of $\alpha$ also leads to second law violations.


Next, we consider a concrete example of a topology changing process involving the merger of two spherical black holes.  Let the horizons be foliated by some parameter $t$ such that $t$ is increasing towards the future on each horizon generator, at $t = -\infty$ the horizon slices have the topology of two spheres, while at $t = +\infty$ the horizon slices have the topology of one sphere.    At some special value of $t$ there is a transition between the two topologies, which occurs at a single point.  The location of this merger point depends on the choice of foliation.  If a classical second law holds analogous to the area increase theorem of General Relativity, the entropy should be increasing no matter what foliation is used.

Comparing the entropy just before and after the merger, the area changes continuously, but the Euler number suffers an instantaneous jump at the exact moment of topology change.  For $\alpha > 0$, this jump decreases the entropy.  Since all other contributions to the entropy change continuously, there is no way to compensate for the instantaneous decrease of the contribution from Gauss-Bonnet term. This leads to an instantaneous violation of the classical second law. This fact was first noticed in Ref. \cite{ted}, and is also discussed in Ref. \cite{Liko}.  On the other hand, for $\alpha < 0$, the second law can be violated when a black hole forms from collapse, at the instant that the horizon first appears.



A similar problem arises for the $m$-th order Lovelock term in a $D = 2m$ dimensional spacetime.  As in the case of Gauss-Bonnet in $D = 4$, any such Lovelock term adds a contribution to the action proportional to the Euler number $\chi$ of the horizon.  Since $\chi = 2$ for any even-dimensional sphere, this can cause an instantaneous entropy decrease at the moment of the black hole with spherical topology merger.  Assuming that the black hole merger happens at a single point, dimensional analysis reveals that the lower-order Lovelock contributions to the entropy evolve continuously through the moment of black hole merger. Therefore, nothing can compensate for the instantaneous decrease of entropy.  Hence, it follows that the second law is violated in any topological Lovelock theory. 


\section{Dimensional Reduction of 5D Gauss-Bonnet}\label{GB}

Now we will utilize these four dimensional results to show a violation of the classical second law in five dimensional Gauss-Bonnet gravity.   Our strategy will be to start with $D = 5$-dimensional spacetime, and compactify the fifth dimension to obtain an effective four dimensional theory.\footnote{Why not simply collide two spherical black holes in $D = 5$?  Because dimensional analysis shows that there is no instantaneous change of the entropy when two 5 dimensional black holes merge at a single point.}  We will then show that this theory violates the classical second law.  Because every solution of the compactified theory corresponds to a solution of the noncompactified theory, and the operation of finding the Wald entropy from a Lagrangian commutes with compactification, the noncompactified theory must also violate the second law.

We will consider a spacetime which is a product of some four dimensional manifold with a circle: $M^{4} \times S^{1}$, with all fields being translation-invariant going around each circle.  This dimensional reduction of higher dimensional Gauss-Bonnet gravity is studied in Ref. \cite{Hoissen}, which showed that the effective four dimensional theory is described by a four-dimensional Einstein-Maxwell-dilaton action with non-minimal coupling terms.  Spacetime indices run from $a, b= \{0,...,\,4\}$. We will set $g_{\alpha D} = 0$ for $\alpha = \{0,...,3\}$, and $g_{44} \equiv \phi$.  This corresponds to a Kaluza-Klein spacetime in which all vector excitations vanish.  (The parity symmetry of the fifth dimension will guarantee that any universe which begins without vector fields will continue to evolve without producing them.)


The Einstein-Gauss-Bonnet Lagrangian can be writen as
\bea
\mathcal{L} = R + \alpha \left (  R^{ab}_{cd} * R^{ef}_{gh}  \right),
\eea
where we are using the notation
\bea
R^{ab}_{cd} * R^{ef}_{gf}  = \frac{1}{2^2}\delta^{cdgh}_{abef} R^{ab}_{cd} R^{ef}_{gh}.
\eea
The dimensional reduction of the five dimensional curvature tensor gives
\bea
^5\!R_{ab}^{cd} = \,^4 \! R_{ab}^{cd} + {}^4 \psi_{ab}^{cd}
\eea
where we have defined
\bea
^n\psi_{ab}^{cd} = - 2 \delta_{4[a} \delta^{4 [c} \phi^{-1} \,^n\del_{b]}{}^n\del^{d]} \phi,
\eea
where $^n \nabla_a$ refers to the covariant derivative operator internal to the n-dimensional space.  Using this, we perform a dimensional reduction of the Gauss-Bonnet Lagrangian:
\bea 
^5\!\!\left(R * R \right) = \,^4 \! \left(R * R + 2R * \psi  + \psi * \psi  \right), \label{compact}
\eea
We also need to perform compactification of the Einstein part of the full action.  The final result for the effective four dimensional action is
\bea
{\cal I} = \frac{1}{16 \pi} \int d^4 x \sqrt{^4 g ~\phi} \,\,^4\! \left(R  + \psi + \alpha \left(R * R  + 2R * \psi  + \psi * \psi \right) \right),
\eea
where $^n \psi = \delta^{ab}_{cd}\psi_{ab}^{cd} =  - 2 \phi^{-1}\,^n\nabla^2 \phi$.  Following the Wald formalism, any black hole solution of this theory will have an entropy given by
\bea
S = \frac{1}{4} \int_{{\cal B}} d^2 x \sqrt{\sigma \phi} \left(1 + 2 \alpha~ ^2\!R + {}^2 \psi \right),
\eea
evaluated on any two dimensional horizon slice ${\cal B}$.  Unlike the case of four dimensions, here we have to take $\alpha > 0$; otherwise the original five dimensional theory will be unstable.  Then the term proportional to the integral of the two dimensional Ricci scalar of the horizon slice will suffer an abrupt decrease, leading to the violation of classical second law, whenever two black holes merge with each other---provided that the term proportional to the integral of $\psi$ also varies smoothly across the merger.

The treatment of the $\psi$ term is somewhat delicate, since it is defined using the derivatives of $\phi$ at the horizon, but the horizon is not typically a differentiable manifold due to nonsmoothness that occurs as a result of horizon generators meeting.  Since black hole generators always meet whenever the topology of the horizon changes, there is the question of how to define $\psi$ at these singular points.

\begin{figure}[ht]
\centering
\includegraphics[width=\textwidth]{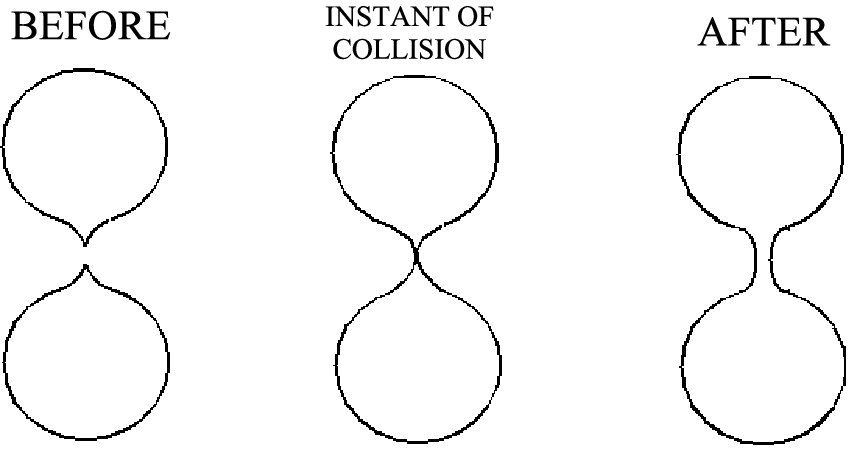}
\caption{\small{The head-on collision of two black holes, pictured on time slices before, at, and after the moment of joinder.  While the horizons may be smooth after the collision, before the collision each of the two black holes must have a nonsmooth point.}}\label{collide}
\end{figure}

To make the question more concrete, let us take a specific example of two equal black holes which collide head on.  This solution is symmetric about rotations around the axis of collision, and is pictured in Fig. \ref{collide}.  As the two black holes begin to merge, they each shoot out horns which end in a nonsmooth point.  These horns then merge at the point of joinder.  The reason for the instantaneous decrease in the entropy at the moment of the collision is that the Ricci scalar $^2\!R$ has a delta function component at the nonsmooth point.  In order to prove that the second law is violated, we need to show that the $\psi$ field does not have a counterbalancing delta function at the nonsmooth part of the horizon.

Since the horizon is non-differentiable, in order to define $\psi$, we will view the horizon as the limit of a smooth (or at least differentiable) surface.  Suppose that the horizon ${\cal H}$ is replaced with another surface ${\cal H^\prime}$ in which the nonsmooth point $X$ and its neighborhood is replaced with a smooth surface $L$ of characteristic length scale $r$.  We can calculate the entropy and then take the limit $r \to 0$.  Although the horizon is nonsmooth, the spacetime $M^4$ in which it is embedded should still be smooth at the horizon.  The scalar fields $\phi$ should also be smooth.  Therefore, one can then apply dimensional analysis to the $\psi$ integral:
\bea
\int d^2 x \sqrt{\sigma \phi}~ \psi = - 2 \int d^2 x \sqrt{\sigma \phi} ~\phi^{-1}\,^2\nabla^2 \phi.
\eea
As the characteristic distance scale $r$ varies over over short distances, the size of the integration measure $d^2 x \sqrt{\sigma}$ should scale like $r^2$.  The integrand is composed of $\phi$ and its derivatives on ${\cal H^\prime}$.  Since ${\cal H^\prime}$ is a smooth submanifold of $M^4$, on which $\phi$ should be smooth, it follows that in the limit as $r \to 0$, one expects the derivatives of the field on $H^\prime$ and $M^4$ to be of the same order:
\begin{eqnarray}
\lim_{r \to 0} \phi(L) = \phi|_X , \\
\lim_{r \to 0} \nabla_a \phi(L) \sim || \nabla_a \phi ||_X , \\
\lim_{r \to 0} \nabla_a \nabla_b  \phi(L) \sim || \nabla_a \nabla_b \phi ||_X,
\end{eqnarray}
and so on.  From this it can be seen that in the $r \to 0$ limit, the integrand converges to a constant value.  Therefore the entire $\psi$ integral has no contribution localized at $X$. Then the instantaneous decrease of the term involving the integral of $^2 \! R$ must lead to an instantaneous violation of the classical second law.

Let us summarize our conclusions.  We started with a five dimensional Gauss-Bonnet theory, and performed a dimensional reduction to get an effective four dimensional action.  Then we showed that any process involving the merger of two black holes results in a decrease of the Wald entropy at least instantaneously.  Since every solution of the compactified theory corresponds to a solution of the original non-compactified theory, these solutions of five dimensional Einstein-Gauss-Bonnet gravity have a violation of the classical second law.  We again stress that this violation has been shown only for one possible way to extend the entropy formula from stationary solutions to this dynamical scenario.  A conservative conclusion might be that the Wald ambiguity should be resolved in some other way.


\section{Dimensional Reduction of Arbitrary Lovelock Theories}\label{arb}


Next, we would like to extend our result to higher order Lovelock theories. For example, the $m = 4$ term would be
\bea
R * R * R * R.
\eea
This term is topological in $2m$ dimensions.  The corresponding Wald entropy is the $(m - 1)$-th Lovelock term evaluated on the horizon.  Therefore, for any Lovelock theory in $(2m + p)$-dimensions whose highest power term is $m$, one can compactify $p$ of the dimensions into a torus in order to end up with a term in the action which is topological.  The $m = 4$ contribution to the compactified entropy takes the form
\bea\label{m3}
S & \propto & \int_{{\cal B}} d^6x \sqrt{^6\sigma ~\phi_1 \phi_2 ...\phi_p} ~ ^6( R * R * R \,+\, 3R * R * \psi  \nonumber  \\ &+& 
3R * \psi * \psi \,+\, \psi * \psi * \psi),
\eea
where each $\psi$ now includes a sum over the internal dimensions of the torus.  As with the Gauss-Bonnet term, we are resolving the Noether charge ambiguity by assuming that the Lovelock entropy can be written entirely in terms of the metric on the horizon.

The first term in Eq. (\ref{m3}) leads to an entropy decline whenever two black holes with spherical topology merge.  To complete the argument, it is necessary to demonstrate that none of the terms which involve $\psi$ can counterbalance this decline as a result of the nonsmooth points on the horizon of the merging black holes.  As before, replace the nonsmooth point with a smooth region with characteristic distance scale $r$.  The integration measure now scales like $r^{(2m-2)}$, each Riemann curvature tensor scales like $r^{-2}$, and the $\phi$ terms scale like $r^0$.  The conclusion is that only the topological Lovelock term can have a contribution from the singular point in the $r \to 0$ limit and the instantaneous violation of the classical second law is inevitable. 

Since the Einstein-Hilbert action is the $m = 1$ Lovelock term, it is worth pointing out why this argument does not apply to it, since the Einstein theory does satisfy the second law. In order to construct a parallel argument for General Relativity, one would compactify all but $2$-dimensions and then collide two 2 dimensional black holes.  But it is impossible for black holes to collide in two dimensions.  For any timelike worldline caught between two colliding black holes would have to fall across one or the other horizon, since no perpendicualar direction is available to escape .  But that would mean that the worldline would already be inside the region of no escape, and thus the zone ``in between'' the black holes ought to have already been included in the black hole interior region. Hence, there really was only one black hole all along!

\section{Conclusion and Open Issues}\label{dis}

The conclusion of this article is that the proposed black hole entropy Eq.~({\ref{m3}) does not always increase during the merger of two black holes.  However, there are a number of different possible inferences that might be drawn.  Most conservatively, it could be that Lovelock black holes do have an increasing entropy, but that entropy is not given by Eq.~({\ref{m3}) but by some other fomula.  Wald and Iyer only derive Eq. ({\ref{m3}) up to ambiguity terms.  These ambiguity terms do not matter for stationary black holes, but they do affect the entropy of nonstationary horizons, as when two black holes merge.  All possible ways to resolve the ambiguity obey the first law of black hole mechanics (which only concerns first order variations away from a stationary solution), but not all choices obey a second law.

However, because the merging black holes are nearly stationary well before and well after the collision, the initial and final values of the entropy are the same no matter how the ambiguities are resolved.  So a different choice of Wald entropy can only remove a temporary entropy decrease, not a permanent entropy decrease.  In the case of nontopological theories such as $D > 4$ Gauss-Bonnet, our compactification argument involves zooming in at the point where the two black holes merge.  It is unclear whether the entropy decrease is permanent or not, so it may be that a different choice of Wald entropy would salvage the second law. However we have not yet been able to find any choice of the entropy which does obey a second law.   

Eq.~(\ref{entropyWald}) by itself, evaluated on a non-stationary horizon, seems like another natural prescription for the entropy.  Because it depends only on the curvature of spacetime, and not to the extrinsic curvature of the horizon, it cannot change discontinuously.  However, it does not work for Einstein-Gauss-Bonnet gravity.  In the case of a $D = 5$ black hole forming from spherically symmetric collapse, the leading order contribution to the initial growth of entropy turns out to be proportional to the time-time component of the Einstein tensor.  But in $D = 5$ Gauss-Bonnet gravity, this component of the Einstein tensor can take either sign.  Consequently this entropy does not increase either.

In the case of topological Lovelock theories, such as $D = 4$ Gauss-Bonnet, there is a permanent decline in the entropy for sufficiently small colliding black holes.\footnote{As pointed out by the referee, this depends on the assumption that the entropy of two infinitely separated black holes is the sum of the entropy of the two individual black holes.}  Therefore, using a different choice of Wald entropy cannot save the topological theories.  In order to salvage the classical second law in this case, we propose that Wald's formula for the entropy is only valid up to the addition of topological terms.  Topological terms in the entropy come from topological terms in the action, but topological terms in the action do not affect the equations of motion.  Since the validity of a classical second law depends only on the equations of motion, it therefore seems that the addition of a topological term to the action ought not to affect the entropy.  Furthermore, the addition of a topological term to the entropy does not affect the validity of the first law, since all first order variations of a horizon preserve the topology, and the first law is only concerned with changes in entropy.  It would therefore seem like one is free to add any topological term to the entropy in order to make it satisfy a second law.  For example, Einstein-Gauss-Bonnet does obey a second law, if one uses the area instead of Eq. ({\ref{m3}).

Another possible way to save the second law is to postuate corrections to the entropy wherever the horizon is nonsmooth.  Before the merger of two black holes, there are always nonsmooth points on the horizon (see Fig. \ref{collide} and Ref. \cite{ellis}).  We have found a violation on the assumption that the entropy of a nonsmooth surface is the same as the entropy of a limit of smooth surfaces.  However, if there are additional corrections to the entropy on a nonsmooth horizon, the additional effect might save the second law.   An analogy might be the singular ``pinch points'' that arise when water droplets change topology; at these points the hydrodynamic approximation breaks down and the dynamics depend on the microscopic degrees of freedom \cite{drops}.  (The simplest alternative prescription is to only include entropy coming from the smooth parts of the horizon.  This would prevent discontinuous changes in the entropy, but it does not change the asymptotic past or future entropy and therefore fails to save the second law, at least in topological theories.)  One way to start investigating this question would be to calculate the divergent corrections to the semiclassical entanglement entropy at nonsmooth horizon points.

It might be that the validity of the second law depends in some important way on quantum effects.  For example, there might be quantum instabilities coming from the formation of numerous low energy, large entropy black holes in the vacuum.  It might be that unless the Lovelock term is higher order in some small parameter (such as $\hbar$ or the string length), there are large nonclassical effects which are present over the timescales on which the second law decreases.

Another possibly relevant issue is that in dynamical situations, (nontopological) Lovelock gravity has different characteristic surfaces for light and gravity, i.e. in curved spacetimes a graviton can travel faster or slower than light \cite{Myers, Arogone}.  Assuming that there exists at least one matter field that propagates along the lightcone, the true causal horizon would be set by whichever of the two fields is moving outwards faster (which varies from location to location).   Hence the causal horizon may be different from the horizon naively obtained from the metric. If this is the case, presumably one should formulate the classical second law using the real causal horizon.  This observation by itself is insufficient to save the second law, but should be kept in mind for any attempts to prove a modified second law.

In view of these conclusions, one must search for an alternative expression for black hole entropy for theories other than General Relativity, which differs from the usual Wald entropy. The most important test for any such proposal depends on the validity of the classical second law.   However, one must also be prepared to accept another possibility: that back hole thermodynamics is invalid when applied to gravity theories other than General Relativity.  (The second law can also be proven for a Wald entropy in $f(R)$ gravity, but any $f(R)$ gravity is conformally equivalent to General Relativity coupled to scalar fields.)  Other theories might be unstable (the positive energy theorem has not been extended to a general Lovelock theory and if such an extension is not possible, the validity of classical second law remains unclear), or unable to arise from any UV complete quantum gravity theory.

\small
\section*{Acknowledgements}
This work was supported by NSF grants PHY-0601800, PHY-0903572 and Maryland Center for Fundamental Physics. We are  grateful for conversations with Ted Jacobson and William Donnelly.


\begin{thebibliography}{100}  

\bibitem{hawking}
S.W. Hawking, ``Particle creation by black holes'', Comm. Math. Phys. {\bf 25}, 152 (1972).

\bibitem{ellis}
S.W. Hawking and G.R.F. Ellis,
\textit{The Large Scale Structure of Spacetime}, Cambridge University Press, 1973.

\bibitem{bekenstein}
J.D. Bekenstein, ``Generalized second law of thermodynamics in black-hole physics'', Phys. Rev. {\bf D9}, 3292(1974).

\bibitem{tedkang}
T. Jacobson, G. Kang, R.C. Myers, ``Increase of Black Hole Entropy in Higher Curvature Gravity'', Phys. Rev. {\bf D52}, 3518 (1995)   
[arXiv:gr-qc/9503020].

\bibitem{wald1}
R.M. Wald,  ``Black Hole Entropy is Noether Charge'', Phys. Rev. {\bf D48}, R3427 (1993) [arXiv:gr-qc/9307038].

\bibitem{wald}
 V. Iyer and R.M. Wald, ``Some properties of Noether charge and a proposal for dynamical black hole entropy,''
Phys.  Rev. {\bf D50}, 846 (1994) [arXiv:gr-qc/9403028].
  
\bibitem{tedon}
T. Jacobson, G. Kang, R.C. Myers, ``On Black Hole Entropy'', Phys. Rev. {\bf D49}, 6587 (1994) [arXiv:gr-qc/9312023].

\bibitem{ted}
T. Jacobson and R.C. Myers, ``Entropy of Lovelock Black Holes'', Phys. Rev. Lett. {\bf 70}, 3684 (1993),
[arXiv:hep-th/9305016]. 

\bibitem{Liko}
T. Liko, ``Topological deformation of isolated horizons'', Phys. Rev. {\bf D77}, 064004 (2008) [arXiv:0705.1518].

\bibitem{lovelock}
C. Lanczos, Z. Phys. {\bf 73}, 147 (1932); Annals Math. {\bf 39},
842 (1938);  D. Lovelock, ``The Einstein Tensor and Its Generalizations'', J. Math. Phys. {\bf 12}, 498
(1971).
 
\bibitem{string1}
B. Zwiebach, ``Curvature Squared Terms and String Theories,'' 
Phys. Lett. {\bf B156}, 315 (1985).


\bibitem{instability}
 D.G. Boulware, S. Deser, ``String-Generated Gravity Models'' Phys. Rev. Lett. {\bf 55}, 2656 (1985).

\bibitem{Hoissen} 
F. M\"uller-Hoissen, ``Non-minimal coupling from dimensional reduction of the Gauss-Bonnet action'', Phys. Lett. {\bf  B201}, 325 (1988). 

\bibitem{drops}
J. Eggers, ``Theory of drop formation'', Physics of Fluids {\bf 7}, 941 (1994) [arXiv:physics/0111003v1].

\bibitem{Myers}
M. Brigante, H. Liu, R.C. Myers, S. Shenker, S. Yaida, ``Viscosity bound violation in higher derivative gravity'', Phys. Rev. {\bf D77}, 126006, (2008) [arXiv:0712.0805].

\bibitem{Arogone}
C. Aragone, \textit{Stringy Characteristics of Effective Gravity}, in SILARG VI : proceedings, edited by M. Novello, WorldScientific, Singapore, (1988); Y. Choquet-Bruhat, J. Math. Phys. {\bf  29}, 1891 (1988).
\end{thebibliography}
\end{document}